\documentclass{natcom}

\usepackage{longtable}
\usepackage{lineno}

\usepackage{epsfig}
\usepackage{color}
\usepackage{bm}
\usepackage{graphicx}
\usepackage{longtable}
\usepackage{amssymb}
\usepackage{amsmath}
\usepackage{rotating}

\newcommand{\apj}{Astrophys. J.}

\newcommand{\pasp}{Publ. Astron. Soc. Pac.}
\newcommand{\apjs}{Astrophys. J. Supp.}
\newcommand{\araa}{Annu. Rev. Astron. Astrophys.}
\newcommand{\mnras}{Mon. Not. R. Astron. Soc.}
\newcommand{\apjl}{Astrophys. J. Let.}
\newcommand{\aap}{Astron. Astrophys.}
\newcommand{\aj}{Astron. J.}
\newcommand{\nat}{Nature}
\newcommand{\na}{New Astron. Rev.}

\title{Carbon monoxide in an extremely metal-poor galaxy}

\author{Yong Shi$^{1,2,3}$, Junzhi Wang$^{4,5}$, Zhi-Yu Zhang$^{6,7}$, Yu Gao$^{8,5}$, Cai-Na Hao$^{9}$, Xiao-Yang Xia$^{9}$, Qiusheng Gu$^{1,2,3}$\\}
\begin{document}

\maketitle

\let\thefootnote\relax\footnote{
\begin{affiliations}

\item School of Astronomy and Space Science, Nanjing University, Nanjing 210093, China.
\item Key Laboratory of Modern Astronomy and Astrophysics (Nanjing University), Ministry of Education, Nanjing 210093, China.
\item Collaborative Innovation Center of Modern Astronomy and Space Exploration, Nanjing 210093, China.
\item Shanghai Astronomical Observatory, Chinese Academy of Sciences, 80 Nandan Road, Shanghai 200030, China.
\item Key Laboratory of Radio Astronomy, Chinese Academy of Sciences, Nanjing 210008, China.
\item Institute for Astronomy, University of Edinburgh, Royal Observatory, Blackford Hill, Edinburgh EH9 3HJ, UK.
\item ESO, Karl-Schwarzschild-Strasse 2, 85748 Garching, Germany
\item Purple Mountain Observatory, Chinese Academy of Sciences, 2 West Beijing Road, Nanjing 210008, China.
\item Tianjin Astrophysics Center, Tianjin Normal University, Tianjin 300387, China. Correspondence and requests for materials should be addressed to Y.S. (email: yshipku@gmail.com).

\end{affiliations}
}
\vspace{-3.5mm}
\begin{abstract}

Extremely metal-poor galaxies with metallicity below 10\% of the solar
value in  the local universe  are the best analogues  to investigating
the interstellar medium at a  quasi-primitive environment in the early
universe.  In  spite  of  the  ongoing formation  of  stars  in  these
galaxies, the presence of molecular gas (which is known to provide the
material reservoir for  star formation in galaxies, such  as our Milky
Way)  remains  unclear.   Here,  we report  the  detection  of  carbon
monoxide (CO), the  primary tracer of molecular gas, in  a galaxy with
7\% solar metallicity,  with additional detections in  two galaxies at
higher metallicities.  Such detections  offer direct evidence  for the
existence of molecular gas in  these galaxies that contain few metals.
Using archived infrared data, it is  shown that the molecular gas mass
per CO luminosity   at  extremely   low  metallicity   is  approximately
one-thousand times the Milky Way value.

\end{abstract}

Galaxies in  the early universe  contain few metals  (elements heavier
than helium) and  dust grains\cite{Walter12}.  On the  surface of dust
grains,  hydrogen   atoms  are   combined  efficiently   into  hydrogen
molecules\cite{Wolfire08}, which serve as the fuel  of star formation in 
present-day spiral galaxies including  our own Milky Way galaxy\cite{Gao04}.
The
lack of metals  thus poses a question regarding the presence of molecular gas
in   the  primordial   galaxies through, e.g.,  the   gas-phase
reaction\cite{Abel97}.  The extremely metal-poor galaxies in the local
universe, with  the oxygen  abundance relative  to hydrogen  below ten
percent   of the solar value,   provide  the   best  local   insights  into
understanding  the  interstellar  medium in  a  quasi-primitive
environment. Although there is  indirect evidence  for the  presence of
molecular  gas in  these galaxies\cite{Madden97,  Hunt10, Shi14}  the
emission from the molecule carbon  monoxide (CO), which is the primary
tracer   of    molecular   gas,    has   never   been    detected   in
them\cite{Bolatto13,Elmegreen13, Warren15,Rubio15,Hunt15,Amorin16}.

In this study, we report the detection of CO in a galaxy at 7\% of solar
metallicity, along with  additional detections in galaxies at  13\% and
18\% solar metallicity; these data offer direct evidence for
the  existence of  molecular  gas  in these  metal-poor galaxies.  By
comparing this data to the gas mass as traced by dust emission, 
the molecular  gas mass per  CO luminosity  in these galaxies  is found to be
much higher than that of the Milky Way galaxy.\\

{\noindent \bf Results}

The galaxy DDO 70  is an extremely metal-poor galaxy  at a distance of
1.38 Mpc  (ref-\citenum{Tully13}), with the gas-phase oxygen  abundance relative
to  hydrogen 12+log(O/H)=7.53  (ref-\citenum{Kniazev05}), compared  to
the solar abundance  at 12+log(O/H)=8.66 (ref-\citenum{Asplund09}). We
have  observed  two additional    dwarf  galaxies  at  somewhat  higher
metallicity,   including   DDO   53   and   DDO   50   at   3.68   Mpc
(ref-\citenum{Tully13})  and  3.27  Mpc  (ref-\citenum{Tully13}),  respectively,  with
12+log(O/H)=7.82       (ref-\citenum{Croxall09})        and       7.92
(ref-\citenum{Croxall09}),  respectively.  As  shown in  Fig. 1,  we
targeted four  dusty star-forming regions  in these three  galaxies as
listed  in  Table 1, using the  Institut  de Radioastronomie  Millimetrique
(IRAM) 30-m  telescope.  For each star-forming region,  we pointed the
telescope to the far-infrared peak that traces gas density enhancement
with ongoing star  formation.  No prior CO detections of these regions
have been reported, possibly because previous works targeted the  peak of
atomic gas  often with a short exposure time\cite{Warren15}.

We  detected CO  $J$=2-1  emission in  all  four star-forming  regions
including one in DDO 70 at 7\%  (labeled as DDO70-A), one in DDO 53 at
14\%  (labeled as  DDO53-A) and  two  in DDO  50 at  18\% (labeled  as
DDO50-A and DDO50-B)  solar metallicity.  The spectra  and results are
shown in  Fig.  1  and listed  in Table  1.  The  1-$\sigma$ continuum
sensitivity is  3.94 mK,  4.17 mK,  3.13 mK and  4.89 mK  for DDO70-A,
DDO53-A, DDO50-A  and DDO50-B, respectively, at  a spectral resolution
of 0.5  km/s, 1.0 km/s, 4.0  km/s and 1.0 km/s,  respectively.  The CO
$J$=2-1 of DDO70-A has a S/N of  5.5 with a full width at half maximum
(FWHM) of 2.4 km/s, and the CO $J$=2-1 of DDO53-A is detected at a S/N
of 7.1 with a FWHM of 7 km/s. The CO line of DDO50-A has an integrated
S/N of 5.9 with a FWHM of  18 km/s. The emission of DDO50-B appears to
have two velocity components.  A single Gaussian fitting gives a value
of S/N  of 6.1 for the  integrated strength peaked at  the velocity of
163 km/s with a  FWHM of 10 km/s, and two  Gaussian fittings give S/Ns
of  6.1 and  3.2 for  the two  components at  161 km/s  and 167  km/s,
respectively, with FWHMs of 3.2  km/s and 3.4 km/s, respectively.  The
CO  $J$=1-0 transition  was covered  by  our observation  but was  not
detected.     The    3-$\sigma$    lower-limits    to    the    ratios
CO(J=2-1)/CO(J=1-0) in the main-beam temperature are 1.9, 0.9, 1.5 and
0.9 for DDO70-A,  DDO53-A, DDO50-A and DDO50-B,  respectively.  As the
size  of  a  CO-emitting  region  shrinks  significantly  at  the  low
metallicity\cite{Rubio15},  we assumed  point sources  for CO-emitting
regions relative to our IRAM beam ($\sim$ 100-200 pc). The above ratio
is thus still  consistent with the assumption that the  CO emission is
thermalized and optically thick.

Fig.  2 shows the total infrared luminosity (8-1000 $\mu$m) versus the
CO luminosity as  well as the star formation rate  (SFR) versus the CO
luminosity  of these  metal-poor star-forming  regions.  Here,  the CO
luminosity defined for  J=1-0 is obtained with  $L'_{\rm CO}$(J=1-0) =
$L'_{\rm CO}$(J=2-1)  by assuming  optically-thick and  thermalized CO
emission.  Both  the infrared  luminosity and  SFR are  measured after
convolution  to match  the beam  of  the IRAM  30m at  the CO  $J$=2-1
frequency.    For  comparison,   massive   star-forming  galaxies   of
approximately   solar  metallicity\cite{Gao04,   Genzel10}  are   also
included in the figure. As is well known, the CO luminosity is related
to  both  far-IR  luminosities  and SFRs  among  massive  star-forming
galaxies, indicating  that the molecular gas  mass as traced by  CO is
related  to  star-formation  activities.   At a  low  metallicity,  CO
decreases because of  not only the eliminated reservoir  of carbon and
oxygen elements but also the increased dissociation of CO molecules by
UV photons under the condition of low dust extinction. As indicated in
the figure,  both infrared/CO and  SFR/CO ratios at a  low metallicity
are significantly  higher than those  of massive galaxies.  In massive
galaxies, IR  luminosity is a good  tracer of the SFR,  accounting for
only a part of  the SFR because of the low  dust content in metal-poor
galaxies.  As a result, the increase  in the SFR/CO ratio from massive
galaxies to  metal-poor ones is  greater than that in  the infrared/CO
ratio.\\

{\noindent \bf Discussions}

The detection of  CO in these objects indicates that  molecular gas is
present at a very low metallicity.  This presence  implies that CO can
still be a tracer  of molecular  gas at a very low  metallicity. To  constrain the
conversion factor  from the  CO luminosity to  the molecular gas  mass, we
estimated the  gas mass through  the dust emission.  All  our galaxies
have multi-band infrared  images available in the archive  of the {\it
  Herschel}  Space Observatory  and HI  gas maps  as observed  by Very
Large  Array   (VLA)\cite{Hunter12}.   We  constructed   the  infrared
spectral energy distribution (SED) of  each region covered by the IRAM
30-m beam  (11$''$) and fitted it with a  dust model\cite{Draine07} to
derive the dust mass (see Methods).   We used the gas-to-dust ratio of
an   extremely    metal-poor   galaxy   (Sextans    A,   7\%   solar
metallicity)\cite{Shi14} by  assuming the  gas-to-dust ratio  equal to
8,000($Z$/0.07)$^{-1.0}$  where  $Z$  is  the  metallicity.  Here,  the
function of the gas-to-dust ratio with the metallicity is suggested by
some observations\cite{Remy-Ruyer14}.  Note that we used the same dust
model  set-up as  that for  Sextans A  to derive  the dust  mass, thus
eliminating the  uncertainty caused by the variation  of the  dust grain
properties.  After subtracting the atomic  gas, the molecular gas mass
is  obtained.   The  derived   molecular  gas  has  a relatively  large
uncertainty that results from the photometric error in the IR SED, the
uncertainty in the dust modeling, the HI gas mass error, and the
error of the gas-to-dust ratio (see Methods for the details).

Fig.  3  shows the  conversion factor  of our  metal-poor star-forming
regions along with those in the literatures\cite{Sandstrom13, Leroy11,
  Israel97, Shi15}, where the molecular gas content is derived through
the spatially resolved dust and HI  gas map.  While previous works are
limited to  the metallicity 12+log(O/H)  $>$ 8.0, our  study indicates
that the  conversion factor  increases rapidly below  this metallicity
limit.  The extremely metal-poor galaxy DDO 70 has a conversion factor
about three orders of magnitude higher than the value of the Milky Way
galaxy.   Another   three  star-forming   regions  at   10-20\%  solar
metallicity have  conversion factors between  $\sim$ 100 and  500. One
difference in  our study  compared to those  at higher  metallicity is
that  we only  targeted the  intense star-formation  peaks.  In  these
regions, the  strong radiation  field may increase  the effects  of CO
dissociation,  thus   biasing  the  conversion  factor   toward  large
values. However,  these regions are  also IR peaks with  more abundant
dust with respective to the rest  of the galaxy; such dust may protect
CO from dissociation.

In spite  of the large  uncertainties, the derived  conversion factors
are still  useful to  differentiate different theoretical  models that
give  a very  large  range  of predictions  at  a  low metallicity  as
illustrated  in Fig.   3.   The empirical  relationship (solid  yellow
line)\cite{Israel97} based  on data  above 12+log(O/H)=8.0 is  a steep
function,  and its  extrapolations at  our metallicity  are consistent
with our  observations.  Among  all theoretical  models, the  one that
invokes      photodissociation       of      CO       and      H$_{2}$
self-shielding\cite{Glover11} matches the  observations including ours
at  a  very low  metallicity.   Other  models either  over-predict  or
significantly   under-predict  the   data  at   the  low   metallicity
end\cite{Narayanan12, Wolfire10, Feldmann12}.\\

{\noindent \bf Methods}

We carried  out the CO $J$=2-1  observation using the IRAM  30 m during
March 22-29,  2016 (Program ID: 168-15,  PI: Y.  Shi) with  a total of
59.5 hrs granted.  The Eight Mixer Receiver with dual polarization and
the Fourier Transform Spectrometers backend were used.  To have a good
baseline for the  spectrum, we adopted the  standard wobbler switching
mode  at a 0.5-Hz beam  throwing with  an offset  of $\pm$120''.   The
pointing and focusing  were set at the beginning of  each run and then
re-calibrated every two hours by  pointing to the bright quasars close
to our targets.  The data reduction  was performed with CLASS in the GILDAS
package. For  each region, we averaged  all scans to obtain  the final
one.   The   effective  on-source   integration  time   including  two
polarization as indicated by CLASS is  1210 min, 413 min, 369 min, and 556
min for  DDO70-A, DDO53-A,  DDO50-A and  DDO 50-B,  respectively, with
system temperatures of 252 K, 223 K, 382 K and 283 K, respectively.

Supplementary Fig. 1 shows  the CO spectra over a  velocity range of
+/- 150  km/s  to  illustrate the goodness of  a long baseline
for our  observations. The HI spectrum  within each IRAM beam  is also
extracted from the HI data  cube\cite{Hunter12} and overlaid on the CO
spectrum as shown in Supplementary Fig. 1. The CO line is within the
HI  velocity range,  although there  are some  offsets in  the central
velocity between the  two that further validates  the reliability of
our CO detections.

{\bf Infrared  SEDs.} The  infrared images  from 70  to 250  $\mu$m as
shown in Supplementary Fig. 2, were retrieved from the archive
of the {\it Herschel} Space  Observatory.  The spatial resolutions are
about  6$''$,   12$''$  and  18$''$   at  70,  160  and   250  $\mu$m,
respectively.      The     data      were     reduced    using    the
unimap\cite{Traficante11}.   The standard  procedure of  the reduction
includes  the  time ordering  of  the  pixels, signal  pre-processing,
glitch removal, drift  removal, making the noise spectrum  and GLS filter,
map making with generalized least square (GLS), post-processing of the
GLS map and  finally the weighted post-processing of the  GLS map. The
mid-infrared images at 3.6, 4.5, 5.8, 8.0 and 24 $\mu$m were retrieved
from the  archive of the {\it  Spitzer} Space Telescope that  is available
through  the  Local  Volume   Legacy  program\cite{Dale09}  with  the
corresponding spatial  resolutions of  1.6'', 1.7'', 1.9'',  2.0'' and
6'', respectively.

To estimate the  dust mass, we  constructed the  IR SED
based on the {\it Spitzer} and {\it Herschel} images. We first checked
the astrometry using field stars and corrected the offsets between the two
telescopes, about 1 arcsec for DDO 70 and DDO 53, and 7 arcsec for DDO
50.  As  the IRAM  beam size  (11$''$) is  relatively small  given the
spatial resolutions  at those IR wavelengths,  the aperture correction
is important.   We used three approaches to derive  the IR SED.
The  first  approach is  to  assume  point  sources for  the  aperture
corrections at  all IR  wavelengths, and then measure  the flux  within the
IRAM beam for each band at native spatial resolutions.  This approach gives the
largest  possible  aperture corrections,  which  could  be treated  as
upperlimits,  given  that  the   star-forming  regions  are  spatially
resolved at {\it Spitzer} 24 $\mu$m  and {\it Herschel} 70 $\mu$m. The
following  two methods  assume that the  star-forming regions  are extended
sources at {\it Spitzer} 24 $\mu$m and {\it Herschel} 70 $\mu$m
to  correct the  flux loss  for {\it  Herschel} 160  and 250
$\mu$m.  For  the second  approach, we  convolved all  infrared images
above  24  $\mu$m  to  the  SPIRE 250  $\mu$m  using  the  convolution
Kernels\cite{Aniano11}  and then measured  the flux  within the  IRAM beam.
These fluxes are  then corrected for the aperture  loss by multiplying
with the ratio  of the flux of  {\it Spitzer} 24 $\mu$m  at its native
resolution to that  at the convolved resolution.  The  third method is
the same  as the second  one but convolves the  24 $\mu$m, 70  and 160-$\mu$m
images to Gaussian 11$''$ beams excluding the 250-$\mu$m image,
and again, this approach corrects the flux loss with  the ratio of
the 24-$\mu$m flux
at  the native  resolution to  that at  the convolved  resolution. The
derived photometry results using three approaches were found to be
within  50\%.  For our discussion, we have adopted the second approach,
which  adopts  images  convolved  to  the  {\it  Herschel}  250  $\mu$m
resolution  and aperture  corrections based  on the  {\it Spitzer}  24
$\mu$m image,  given that the star-forming regions  are spatially resolved
at 24  $\mu$m.  As we  pointed at the  bright infrared peaks,  the photon
noise is small while the photometric error is instead dominated by the
aperture  correction because of the  small IRAM  beam.  We  assigned 50\%
of fluxes as systematic  uncertainties for  the IR  photometry
at  all wavelengths. The final SED is shown in Supplementary Fig. 3.

{\bf Measurements of dust masses, molecular gas mass, SFRs and stellar
  masses.}   Following our  previous work\cite{Shi14},  the IR  SED is
fitted  with  a dust  model\cite{Draine07}  to  obtain the  dust  mass
measurement.  We adopted  the Milky Way dust grains and  fixed the PAH
fraction to be the minimum given  the low metallicity of our galaxies.
The maximum intensity of the  stellar radiation field is further fixed
to be 10$^{6}$.   Thus, the model has three  free parameters including
the dust mass, the minimum stellar light intensity and the fraction of
dust exposed  to the  minimum radiation field.   An additional  4000 K
black-body  is  included  to  model  the  emission  from  the  stellar
photosphere.  As  shown   in  Supplementary  Fig.  3   and  listed  in
Supplementary Table  1, the fitting  results are reasonably  good.  If
using the SMC dust model, the dust  mass differs by no more than 15\%,
which is consistent with previous studies\cite{Shi14, Draine07b}.

To derive the total gas mass from the dust mass, the gas-to-dust ratio
is needed.   Unlike our previous work\cite{Shi14},  the IR observation
is  not deep  enough  to derive  the gas-to-dust  ratio  based on  the
diffuse  light   for  individual   galaxies.   We  thus   adopted  the
gas-to-dust ratio (8000)  of Sextans A at 7\% solar  from the previous
work\cite{Shi14} that is  based on the diffuse light as  the value for
DDO 70, which is at the same metallicity. Given the large variation in
the gas-to-dust ratio from object to object\cite{Remy-Ruyer14} at this
metallicity,  we assigned  0.5 dex  as 1-$\sigma$  uncertainty of  the
ratio.  For  DDO 50 and  DDO 53, we assumed  a linear increase  of the
gas-to-dust  ratio  with  the  decreasing  metallicity  following  the
literature  study\cite{Remy-Ruyer14}, with a 1-$\sigma$ uncertainty  of
$\sim$ 0.3 dex  at their metallicities.  As discussed  in the previous
work\cite{Shi14},  although different dust  grain models  provide
different dust  masses,  these different dust masses do  not  affect
the  derived  gas masses  because the dust-to-gas ratio changes accordingly.

To obtain  the molecular gas, the  atomic gas mass is  subtracted from
the dust-based total gas mass.  The HI gas maps of three galaxies were
observed  with the  Very  Large  Array through  the  program of  Local
Irregulars That Trace  Luminosity Extremes\cite{Hunter12}.  We adopted
the  robust-weighted   maps  with   the  synthesized  beam   sizes  of
13.8$''$$\times$13.2$''$,          6.3$''$$\times$5.7$''$          and
7.0$''$$\times$6.1$''$ for  DDO 70, DDO  53 and DDO  50, respectively.
Although the  DDO 70  has a  resolution that is slightly  worse than that of
our IRAM beam, the HI emission  is pretty diffuse so that we  can assume the HI
mass  surface   density  measured   at  its   resolution  is   a  good
approximation of that  within the IRAM beam.  For DDO  53 and DDO 50,
we convolved the HI maps to the 11$''$  beam to measure the HI mass, which
is  almost  the  same  ($<$10\%)  as  those  measured  at  the  native
resolution, given  the HI emission  is diffuse.  We also  retrieved the
reduced far-ultraviolet images from the {\it GALEX} data archive whose
spatial    resolution   is    about   5$''$.     The   SFR    is   the
combination\cite{Leroy08} of  the unobscured part (as  traced by far-UV)
and the  obscured part (as traced  by 24 $\mu$m emission).   The SFRs of
massive galaxies used  for comparison in Fig. 2 are  based on their IR
luminosities\cite{Gao04}  using   the  formula\cite{Kennicutt98}  with
corrections for Chabier IMF.  The stellar mass is derived based on the
3.6 and 4.5 $\mu$m emission \cite{Eskew12}.

{\bf Data availability.}   The data that support the  findings of this
study  are available  from  the corresponding  author upon  reasonable
request.

\begin{addendum}
\item [Acknowledgements]

Based on observations carried out under project number 168-15 with the
IRAM  30m Telescope.   IRAM is  supported by  INSU/CNRS (France),  MPG
(Germany) and IGN  (Spain).  Y.S.  acknowledges support  for this work
from  the National  Natural Science  Foundation of  China (NSFC  grant
11373021), the  Strategic Priority  Research Program The  Emergence of
Cosmological Structures of the Chinese  Academy of Sciences (grant No.
XDB09000000),  and  the  Excellent  Youth Foundation  of  the  Jiangsu
Scientific  Committee (grant  BK20150014). J.W.   is supported  by the
National  973 program  (grant  2012CB821805) and  by  the NSFC  (grant
11590783).  Z.Y.Z.   acknowledges support  from the  European Research
Council in  the form of  the Advanced Investigator  Programme, 321302,
COSMICISM. Y.G. is supported by  Pilot-b (XDB09000000) and NSFC (grant
11390373 and 11420101002). C.N.H.   and X.Y.X. acknowledge the support
from the NSFC grant 11373027. Q.G. was supported by the NSFC (11273015
and 11133001) and by the National 973 programme (grant 2013CB834905).

\item[Author  Contributions] 

  Y.S. led the IRAM proposal and the writing of the manuscript. Y.S. and J.W.
  performed the observations and reduced the data. All others helped develop the proposal
  and commented on the manuscript.

\item[Additional information]
  
{\bf Supplementary Information} accompanies this paper at http://www.nature.com/naturecommunications

{\bf Competing financial interests:} The authors declare no competing financial interests.

{\bf Reprints and permission} information is available online at http://npg.nature.com/
reprintsandpermissions/

\end{addendum}

\clearpage

\begin{figure*}[tbh]
\begin{center}
  \includegraphics[scale=0.8]{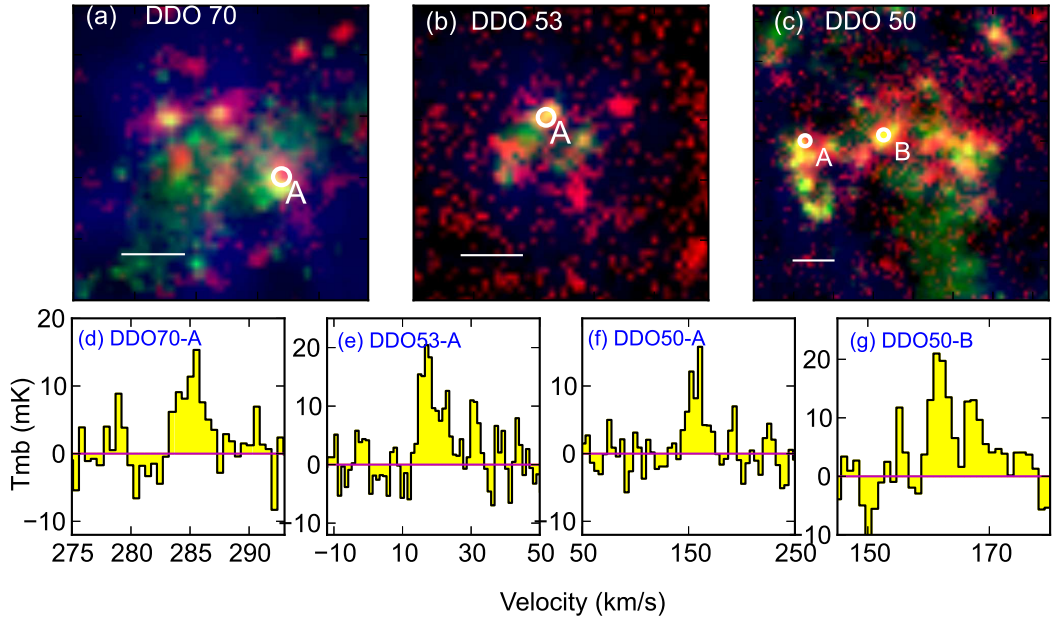}
  \caption{ {\noindent  False-color  images of three  galaxies along
  with the  CO $J$=2-1 spectra}.  (a)  The image of DDO  70, where red
denotes  infrared emission  at 160  $\mu$m, green  denotes the  far-UV
emission, and  blue denotes the  atomic hydrogen 21 cm  emission.  (b)
The image  of DDO 53.  (c)  The image of  DDO 50. All the  white scale
bars are 40'' across.  (d) The CO  $J$=2-1 spectra for region A in DDO
70.  (e) The  CO $J$=2-1 spectra for  region A in DDO 53.   (f) The CO
$J$=2-1 spectra  for region A in  DDO 50.  (g) The  CO $J$=2-1 spectra
for region B  in DDO 50. The  spectral channel width is  0.5 km/s, 1.0
km/s, 4.0 km/s and 1.0 km/s for DDO70-A, DDO53-A, DDO50-A and DDO50-B,
respectively, and  the corresponding 1-$\sigma$  continuum sensitivity
is 3.94 mK, 4.17 mK, 3.13 mK and 4.89 mK, respectively. }
\end{center}
\end{figure*}

\begin{figure*}[tbh]
\begin{center}
  \includegraphics[scale=0.6]{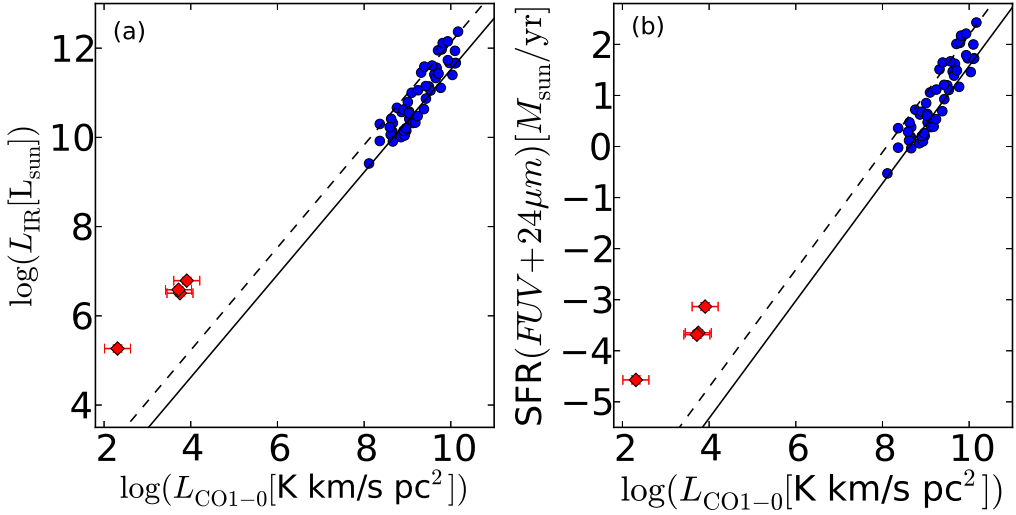}
  \caption{ {\noindent      The    infrared   luminosity   and
  star-formation-rate   vs.   CO   luminosity}.   (a)   The  infrared
luminosity vs.   CO luminosity of our  metal-poor star-forming regions
 compared  to massive star-forming  galaxies.  The error bar  is the
standard  deviation, which  is  basically the  photon  noise for  each
measurement of luminosity.   (b)  The  SFR  vs.   CO  luminosity  of  our  metal-poor
star-forming regions as compared to massive star-forming galaxies. The
error bar is the standard deviation. The error of the CO luminosity is
the photon noise, and the error of  the SFR is the photon noise plus
systematic  uncertainty.   The  red  diamonds denote  our  observed  four
regions,  and  the  blue  circles denote massive  star-forming  galaxies
\cite{Gao04}.  The  solid line  is the  best fit to  star-forming disk
galaxies \cite{Genzel10}, whereas the dashed  line is the best fit to
the star-forming starburst galaxies \cite{Genzel10}.}
\end{center}
\end{figure*}

\begin{figure*}[tbh]
\begin{center}
  \includegraphics[scale=0.8]{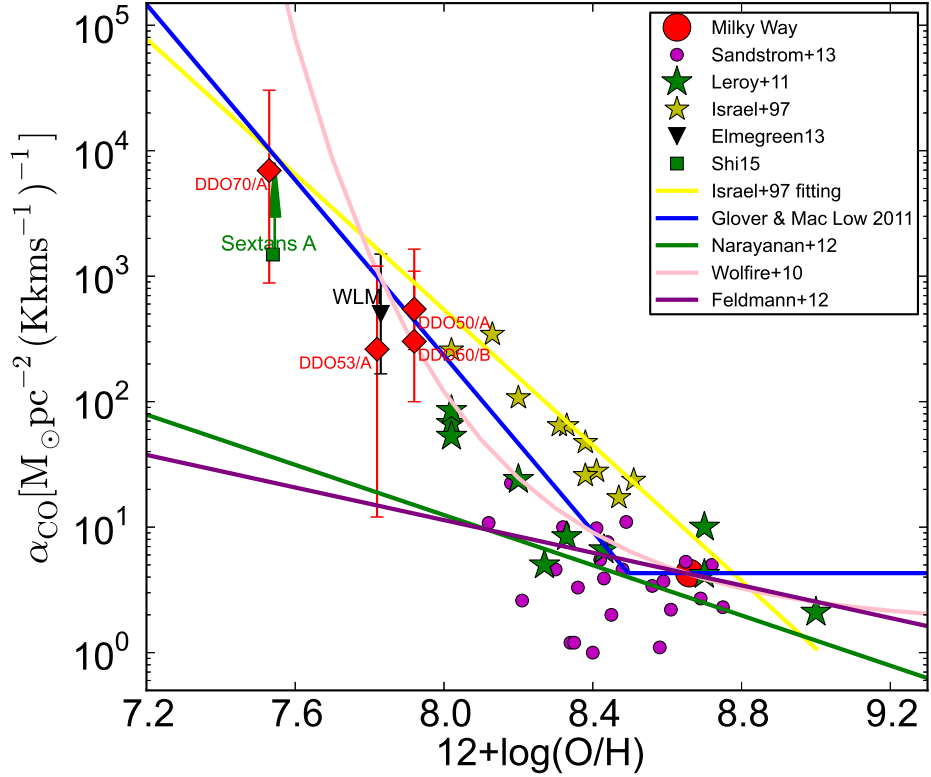}
  \caption{{\noindent The  conversion factor from  CO luminosity
  to molecular gas mass.}  The four red diamonds denote the result of
our metal poor  star-forming regions, and all other  symbols denote those
observations in  the literature \cite{Israel97,  Leroy11, Elmegreen13,
  Sandstrom13}.  The  error bars of  our measurements are  the standard
deviations, which are caused by the uncertainties on the CO luminosity,
the HI gas mass, the dust mass  as well as the dust-to-gas ratio.  The
lines denote  models'   predictions,   including   the  empirical   one
\cite{Israel97} as well as theoretical ones \cite{Wolfire10, Glover11,
  Feldmann12, Narayanan12}.   The parameters in those  models were set
basically following  the literature work\cite{Bolatto13} including a
linear scaling  of the dust-to-gas  ratio with the metallicity,  and a
typical gas surface  density of 100 M$_{\odot}$/pc$^{2}$  in the Milky
Way.}
\end{center}
\end{figure*}


\begin{table*}[b]
\begin{center}{ \bf Table 1. The Properties of IRAM-30m-targeted regions}\end{center}
\tiny
\begin{center}
\begin{tabular}{llllllllllllll}
\hline
name    & D    & 12+log(O/H) & Pos. (J2000)              & $M_{\rm star}$\textdaggerdbl & $L_{\rm 8-1000{\mu}m}$  & SFR\textdaggerdbl             & $T_{\rm peak, CO}$ &  $v_{\rm CO}$ & FWHM$_{\rm CO}$     & $S_{\rm CO}{\Delta}$V  & $L_{\rm CO}$    &  $\alpha_{\rm CO}$  \\
        & (Mpc)&             &                           &   (M$_{\odot}$)              &  L$_{\odot}$  & (M$_{\odot}$/yr)                     & (mK)             & (km/s)       & (km/s)   & (mK km/s)          & (K km/s pc$^{2}$) & (M$_{\odot}$/pc$^{2}$/(K km/s))   \\
\hline                                                            
DDO70-A & 1.38 & 7.53 &  09 59 58.08  +05 19 45.5 &  1.7x10$^{5}$ & (1.8$\pm$0.9)$\times$10$^{5}$ & 2.7x10$^{-5}$ & 13.0            & 285.0           & 2.4$\pm$0.5     & 33$\pm$6      & 204$\pm$37    & 6949$^{+23403}_{-6067}$\\
DDO53-A & 3.68 & 7.82 &  08 34 07.63  +66 10 52.4 &  3.5x10$^{5}$ & (3.2$\pm$1.6)$\times$10$^{6}$ & 2.3x10$^{-4}$ & 16.6            & 17.6            & 7.3$\pm$1.3     & 129$\pm$18    & 5635$\pm$788  & 261$^{+940}_{-249}$    \\
DDO50-A & 3.27 & 7.92 &  08 19 12.30  +70 43 08.3 &  1.1x10$^{6}$ & (6.1$\pm$3.1)$\times$10$^{6}$ & 7.3x10$^{-4}$ & 12.3            & 155.3           & 18.0$\pm$3.5    & 234$\pm$40    & 8084$\pm$1382 & 546$^{+1095}_{-286}$   \\
DDO50-B & 3.27 & 7.92 &  08 19 28.21  +70 43 02.3 &  4.5x10$^{5}$ & (3.8$\pm$1.9)$\times$10$^{6}$ & 2.1x10$^{-4}$ & 13.3            & 163.4           & 10.6$\pm$2.1    & 151$\pm$25    & 5217$\pm$864  & 302$^{+793}_{-202}$    \\
\hline
\end{tabular}
\end{center}
\textdaggerdbl The stellar mass and SFR are measured within the IRAM 30 m beam (11$''$). The uncertainties on both measurements
are dominated by the systematic errors (about 0.3 dex).\\
\end{table*}


\begin{thebibliography}{10}
\expandafter\ifx\csname url\endcsname\relax
  \def\url#1{\texttt{#1}}\fi
\expandafter\ifx\csname urlprefix\endcsname\relax\def\urlprefix{URL }\fi
\providecommand{\bibinfo}[2]{#2}
\providecommand{\eprint}[2][]{\url{#2}}




\bibitem{Walter12}
\bibinfo{author}{{Walter}, F.} \emph{et~al}   
\newblock \bibinfo{title}{Evidence for low extinction in actively star-forming galaxies at z $>$ 6.5}.  
\newblock \emph{\bibinfo{journal}{\apj}} \textbf{\bibinfo{volume}{752}},
  \bibinfo{pages}{93-98} (\bibinfo{year}{2012}).

\bibitem{Wolfire08}
\bibinfo{author}{{Wolfire}, M. G.} \emph{et~al}   
\newblock \bibinfo{title}{Chemical Rates on Small Grains and PAHs: C+ Recombination and H2 Formation}.  
\newblock \emph{\bibinfo{journal}{\apj}} \textbf{\bibinfo{volume}{680}},
  \bibinfo{pages}{384-397} (\bibinfo{year}{2008}).

\bibitem{Gao04}
\bibinfo{author}{{Gao}, Y. and {Solomon}, P.~M.}
\newblock \bibinfo{title}{HCN Survey of Normal Spiral, Infrared-luminous, and Ultraluminous Galaxies}
\newblock \emph{\bibinfo{journal}{\apjs}} \textbf{\bibinfo{volume}{152}},
  \bibinfo{pages}{63-80} (\bibinfo{year}{2004}).

\bibitem{Abel97}
\bibinfo{author}{{Abel}, T.} \emph{et~al}   
\newblock \bibinfo{title}{Modeling primordial gas in numerical cosmology}.  
\newblock \emph{\bibinfo{journal}{\na}} \textbf{\bibinfo{volume}{2}},
  \bibinfo{pages}{181-207} (\bibinfo{year}{1997}).

\bibitem{Madden97}
\bibinfo{author}{{Madden}, S. C.} \emph{et~al}   
\newblock \bibinfo{title}{[C II] 158 Micron Observations of IC 10: Evidence for Hidden Molecular Hydrogen in Irregular Galaxies}.  
\newblock \emph{\bibinfo{journal}{\apj}} \textbf{\bibinfo{volume}{483}},
  \bibinfo{pages}{200-209} (\bibinfo{year}{1997}).

\bibitem{Hunt10}
\bibinfo{author}{{Hunt}, L.} \emph{et~al}   
\newblock \bibinfo{title}{The Spitzer View of Low-Metallicity Star Formation. III. Fine-Structure Lines, Aromatic Features, and Molecules}.  
\newblock \emph{\bibinfo{journal}{\apj}} \textbf{\bibinfo{volume}{712}},
  \bibinfo{pages}{164-187} (\bibinfo{year}{2010}).
  
\bibitem{Shi14}
\bibinfo{author}{{Shi}, Y.} \emph{et~al}   
\newblock \bibinfo{title}{Inefficient star formation in extremely metal poor galaxies}.  
\newblock \emph{\bibinfo{journal}{\nat}} \textbf{\bibinfo{volume}{514}},
  \bibinfo{pages}{335-338} (\bibinfo{year}{2014}).


  
\bibitem{Bolatto13}
\bibinfo{author}{{Bolatto}, A.} \emph{et~al.}
\newblock \bibinfo{title}{{The CO-to-H2 Conversion Factor}}.
\newblock \emph{\bibinfo{journal}{\araa}} \textbf{\bibinfo{volume}{51}},
  \bibinfo{pages}{207-268} (\bibinfo{year}{2013}).

\bibitem{Elmegreen13}
\bibinfo{author}{{Elmegreen}, B.~G. } \emph{et~al}   
\newblock \bibinfo{title}{Carbon monoxide in clouds at low metallicity in the dwarf irregular galaxy WLM}.  
\newblock \emph{\bibinfo{journal}{\nat}} \textbf{\bibinfo{volume}{495}},
  \bibinfo{pages}{487-489} (\bibinfo{year}{2013}).

  
\bibitem{Warren15}
\bibinfo{author}{{Warren}, S.~R.} \emph{et~al}   
\newblock \bibinfo{title}{CARMA CO Observations of Three Extremely Metal-poor, Star-forming Galaxies}.  
\newblock \emph{\bibinfo{journal}{\apj}} \textbf{\bibinfo{volume}{814}},
  \bibinfo{pages}{30-38} (\bibinfo{year}{}).


\bibitem{Rubio15}
\bibinfo{author}{{Rubio}, M.} \emph{et~al}   
\newblock \bibinfo{title}{Dense cloud cores revealed by CO in the low metallicity dwarf galaxy WLM}.  
\newblock \emph{\bibinfo{journal}{\nat}} \textbf{\bibinfo{volume}{525}},
  \bibinfo{pages}{218-221} (\bibinfo{year}{2015}).

\bibitem{Hunt15}
\bibinfo{author}{{Hunt}, L.~K.} \emph{et~al}   
\newblock \bibinfo{title}{Molecular depletion times and the CO-to-H$_{2}$ conversion factor in metal-poor galaxies}.  
\newblock \emph{\bibinfo{journal}{\aap}} \textbf{\bibinfo{volume}{583}},
  \bibinfo{pages}{A114} (\bibinfo{year}{2015}).


\bibitem{Amorin16}
\bibinfo{author}{{Amor{\'{\i}}n}, R.} \emph{et~al}   
\newblock \bibinfo{title}{Molecular gas in low-metallicity starburst galaxies:. Scaling relations and the CO-to-H$_{2}$ conversion factor}.  
\newblock \emph{\bibinfo{journal}{\aap}} \textbf{\bibinfo{volume}{588}},
  \bibinfo{pages}{A23} (\bibinfo{year}{2016}).

\bibitem{Tully13}
\bibinfo{author}{{Tully}, R.~B.} \emph{et~al}   
\newblock \bibinfo{title}{Cosmicflows-2: The Data}.  
\newblock \emph{\bibinfo{journal}{\aj}} \textbf{\bibinfo{volume}{146}},
  \bibinfo{pages}{86-110} (\bibinfo{year}{2013}).

\bibitem{Kniazev05}
\bibinfo{author}{{Kniazev}, A.~Y.} \emph{et~al}   
\newblock \bibinfo{title}{Spectrophotometry of Sextans A and B: Chemical Abundances of H II Regions and Planetary Nebulae}.  
\newblock \emph{\bibinfo{journal}{\aj}} \textbf{\bibinfo{volume}{130}},
  \bibinfo{pages}{1558-1573} (\bibinfo{year}{2005}).

\bibitem{Asplund09}
\bibinfo{author}{{Asplund}, M.} \emph{et~al}   
\newblock \bibinfo{title}{The Chemical Composition of the Sun}.  
\newblock \emph{\bibinfo{journal}{\araa}} \textbf{\bibinfo{volume}{47}},
  \bibinfo{pages}{481-522} (\bibinfo{year}{2009}).

\bibitem{Croxall09}
\bibinfo{author}{{Croxall}, K.~V.} \emph{et~al}   
\newblock \bibinfo{title}{Chemical Abundances of Seven Irregular and Three Tidal Dwarf Galaxies in the M81 Group}.  
\newblock \emph{\bibinfo{journal}{\apj}} \textbf{\bibinfo{volume}{705}},
  \bibinfo{pages}{723-738} (\bibinfo{year}{2009}).

\bibitem{Genzel10}
\bibinfo{author}{{Genzel}, R.} \emph{et~al}   
\newblock \bibinfo{title}{A study of the gas-star formation relation over cosmic time}.  
\newblock \emph{\bibinfo{journal}{\mnras}} \textbf{\bibinfo{volume}{2010}},
  \bibinfo{pages}{2091-2108} (\bibinfo{year}{2010}).

\bibitem{Hunter12}
\bibinfo{author}{{Hunter}, D.~A.} \emph{et~al}   
\newblock \bibinfo{title}{Little Things}.  
\newblock \emph{\bibinfo{journal}{\aj}} \textbf{\bibinfo{volume}{144}},
  \bibinfo{pages}{134-162} (\bibinfo{year}{2012}).

\bibitem{Draine07}
\bibinfo{author}{{Draine}, B.~T. and {Li}, A.} \emph{et~al}   
\newblock \bibinfo{title}{Infrared Emission from Interstellar Dust. IV. The Silicate-Graphite-PAH Model in the Post-Spitzer Era}.  
\newblock \emph{\bibinfo{journal}{\apj}} \textbf{\bibinfo{volume}{657}},
  \bibinfo{pages}{810-837} (\bibinfo{year}{2007}).

\bibitem{Remy-Ruyer14}
\bibinfo{author}{{R{\'e}my-Ruyer}, A.} \emph{et~al}   
\newblock \bibinfo{title}{Gas-to-dust mass ratios in local galaxies over a 2 dex metallicity range}.  
\newblock \emph{\bibinfo{journal}{\aap}} \textbf{\bibinfo{volume}{2014}},
  \bibinfo{pages}{A31} (\bibinfo{year}{2014}).

\bibitem{Sandstrom13}
\bibinfo{author}{{Sandstrom}, K.~M.} \emph{et~al}   
\newblock \bibinfo{title}{The CO-to-H$_{2}$ Conversion Factor and Dust-to-gas Ratio on Kiloparsec Scales in Nearby Galaxies}.  
\newblock \emph{\bibinfo{journal}{\apj}} \textbf{\bibinfo{volume}{777}},
  \bibinfo{pages}{5-37} (\bibinfo{year}{2013}).

\bibitem{Leroy11}
\bibinfo{author}{{Leroy}, A.~K.} \emph{et~al}   
\newblock \bibinfo{title}{The CO-to-H$_{2}$ Conversion Factor from Infrared Dust Emission across the Local Group}.  
\newblock \emph{\bibinfo{journal}{\apj}} \textbf{\bibinfo{volume}{737}},
  \bibinfo{pages}{12-24} (\bibinfo{year}{2011}).

\bibitem{Israel97}
\bibinfo{author}{{Israel}, F.~P.}  
\newblock \bibinfo{title}{H\_2 and its relation to CO in the LMC and other magellanic irregular galaxies}.  
\newblock \emph{\bibinfo{journal}{\aap}} \textbf{\bibinfo{volume}{328}},
  \bibinfo{pages}{471-482} (\bibinfo{year}{1997}).

\bibitem{Shi15}
\bibinfo{author}{{Shi}, Y.} \emph{et~al}   
\newblock \bibinfo{title}{The Weak Carbon Monoxide Emission in an Extremely Metal-poor Galaxy, Sextans }.  
\newblock \emph{\bibinfo{journal}{\apjl}} \textbf{\bibinfo{volume}{804}},
  \bibinfo{pages}{11-14} (\bibinfo{year}{2015}).

\bibitem{Glover11}
\bibinfo{author}{{Glover}, S.~C.~O. and {Mac Low}, M.-M.} 
\newblock \bibinfo{title}{On the relationship between molecular hydrogen and carbon monoxide abundances in molecular clouds}.  
\newblock \emph{\bibinfo{journal}{\mnras}} \textbf{\bibinfo{volume}{412}},
  \bibinfo{pages}{337-350} (\bibinfo{year}{2011}).

\bibitem{Narayanan12}
\bibinfo{author}{{Narayanan}, D.} \emph{et~al}   
\newblock \bibinfo{title}{A general model for the CO-H$_{2}$ conversion factor in galaxies with applications to the star formation law}.  
\newblock \emph{\bibinfo{journal}{\mnras}} \textbf{\bibinfo{volume}{2012}},
  \bibinfo{pages}{3127-3146} (\bibinfo{year}{2012}).

\bibitem{Wolfire10}
\bibinfo{author}{{Wolfire}, M.~G.} \emph{et~al}   
\newblock \bibinfo{title}{The Dark Molecular Gas}.  
\newblock \emph{\bibinfo{journal}{\apj}} \textbf{\bibinfo{volume}{716}},
  \bibinfo{pages}{1191-1207} (\bibinfo{year}{2010}).

\bibitem{Feldmann12}
\bibinfo{author}{{Feldmann}, R. and {Gnedin}, N.~Y. and {Kravtsov}, A.~V.} 
\newblock \bibinfo{title}{The X-factor in Galaxies. I. Dependence on Environment and Scale}.  
\newblock \emph{\bibinfo{journal}{\apj}} \textbf{\bibinfo{volume}{747}},
  \bibinfo{pages}{124-144} (\bibinfo{year}{2012}).



\bibitem{Traficante11}
\bibinfo{author}{{Traficante}, A.} \emph{et~al.}
\newblock \bibinfo{title}{{Data  reduction pipeline   for  the  Hi-GAL   survey}}.
\newblock \emph{\bibinfo{journal}{Maris}} \textbf{\bibinfo{volume}{416}},
  \bibinfo{pages}{2932-2943} (\bibinfo{year}{2011}).

\bibitem{Dale09}
\bibinfo{author}{{Dale}, D.~A.} \emph{et~al}   
\newblock \bibinfo{title}{The Spitzer Local Volume Legacy: Survey Description and Infrared Photometry}.  
\newblock \emph{\bibinfo{journal}{\apj}} \textbf{\bibinfo{volume}{703}},
  \bibinfo{pages}{517-556} (\bibinfo{year}{2009}).



\bibitem{Aniano11}
\bibinfo{author}{{Aniano}, G.} \emph{et~al}   
\newblock \bibinfo{title}{Common-Resolution Convolution Kernels for Space- and Ground-Based Telescopes}.  
\newblock \emph{\bibinfo{journal}{\pasp}} \textbf{\bibinfo{volume}{123}},
  \bibinfo{pages}{1218-1236} (\bibinfo{year}{2011}).


\bibitem{Draine07b}
\bibinfo{author}{{Draine}, B.} \emph{et~al}   
\newblock \bibinfo{title}{Dust Masses, PAH Abundances, and Starlight Intensities in the SINGS Galaxy Sample}.  
\newblock \emph{\bibinfo{journal}{\apj}} \textbf{\bibinfo{volume}{663}},
  \bibinfo{pages}{866-894} (\bibinfo{year}{2007b}).


\bibitem{Leroy08}
\bibinfo{author}{{Leroy}, A.~K.} \emph{et~al}   
\newblock \bibinfo{title}{The Star Formation Efficiency in Nearby Galaxies: Measuring Where Gas Forms Stars Effectively}.  
\newblock \emph{\bibinfo{journal}{\aj}} \textbf{\bibinfo{volume}{136}},
  \bibinfo{pages}{2782-2845} (\bibinfo{year}{2012}).

\bibitem{Kennicutt98}
\bibinfo{author}{{Kennicutt}, R.}   
\newblock \bibinfo{title}{Star Formation in Galaxies Along the Hubble Sequence}.  
\newblock \emph{\bibinfo{journal}{\araa}} \textbf{\bibinfo{volume}{36}},
  \bibinfo{pages}{189-232} (\bibinfo{year}{1998}).


\bibitem{Eskew12}
\bibinfo{author}{{Eskew}, M. and {Zaritsky}, D. and {Meidt}, S.} 
\newblock \bibinfo{title}{Converting from 3.6 and 4.5 {$\mu$}m Fluxes to Stellar Mass}.  
\newblock \emph{\bibinfo{journal}{\aj}} \textbf{\bibinfo{volume}{143}},
  \bibinfo{pages}{139-145} (\bibinfo{year}{2012}).


  
\end{thebibliography}
\end{document}